\documentclass[runningheads]{llncs}
\usepackage{longtable}
\usepackage{makecell}
\usepackage{adjustbox}
\usepackage{graphicx}
\usepackage{multirow}
\usepackage[super]{nth}
\usepackage{xr}
\begin{document}
\title{Generative AI for Rapid Diffusion MRI with Improved Image Quality, Reliability and Generalizability}
\titlerunning{Generative AI for Rapid dMRI}
%
\author{Amir Sadikov\inst{1}\inst{2} \and Xinlei Pan\inst{3} \and Hannah Choi\inst{1} \and Lanya T. Cai\inst{1} \and Pratik Mukherjee\inst{1}\inst{2}}
\authorrunning{Sadikov et al.}
%
\institute{Radiology and Biomedical Imaging, University of California, San Francisco \and Graduate Group in Bioengineering, University of California, San Francisco \and University of California, Berkeley\\
\email{amir.sadikov@ucsf.edu}}
%
%
\maketitle              
\begin{center}\textbf{Abstract}\end{center}

\noindent\textbf{Background}
Diffusion MRI (dMRI) is a non-invasive, in-vivo biomedical imaging method for mapping tissue microstructure. Applications include structural connectivity imaging of the human brain, as well as detecting microstructural neural changes not detectable by other clinical neuroimaging techniques. However, acquiring high signal-to-noise ratio (SNR) dMRI datasets with high angular and spatial resolution requires prohibitively long scan times, limiting usage in many important clinical settings, especially for children, the elderly, and in acute neurological disorders that may require conscious sedation or general anesthesia.

\noindent\textbf{Methods}
We employ a Swin UNEt Transformers (Swin UNETR) model, trained on augmented Human Connectome Project (HCP) data and conditioned on registered T1 scans, to perform generalized denoising of dMRI. We also qualitatively demonstrate super-resolution with artificially downsampled HCP data in normal adult volunteers.

\noindent\textbf{Results}
Remarkably, Swin UNETR can be fine-tuned for an out-of-domain dataset with a single example scan, as we demonstrate on dMRI of children with neurodevelopmental disorders and of adults with acute evolving traumatic brain injury, each cohort scanned on different models of scanners with different dMRI imaging protocols at different sites. This robustness to scan acquisition parameters, patient populations, scanner types, and sites eliminates the advantages of self-supervised machine learning methods over our fully supervised generative AI approach. We exceed current state-of-the-art denoising methods in accuracy and test-retest reliability of rapid diffusion tensor imaging (DTI) requiring only 90 seconds of scan time. Applied to tissue microstructural modeling of dMRI at high diffusion-weighting and therefore low SNR, Swin UNETR denoising achieves dramatic improvements over the state-of-the-art for test-retest reliability of intracellular volume fraction and free water fraction measurements and is able to remove heavy-tail noise, improving biophysical modeling fidelity.

\noindent\textbf{Conclusion} Swin UNeTR enables rapid diffusion MRI with unprecedented accuracy and reliability, especially at high diffusion-weighting for probing biological tissues at microscopic spatial scales for scientific and clinical applications. The code and model are publicly available at \url{https://github.com/ucsfncl/dmri-swin}.
\section{Introduction}
Diffusion MRI (dMRI) can provide valuable clinical information and assess tissue microstructure; however, its low signal-to-noise (SNR) ratio can result in poor diagnostic and quantitative accuracy \cite{Jones2012}. To improve SNR, most dMRI protocols require low angular and spatial resolution or a long scan time, which limits usage in many important clinical settings. Therefore, there is great interest in having short patient scan times without compromising SNR or spatial and angular resolution.

Several supervised methods have been proposed to denoise brain dMRI scans; however, they are limited by their lack of generalizability. Often, they work on only one b-value and a prespecified set of diffusion-encoding directions, are built to predict only one set of microstructural parameters, or are trained and validated on the same dataset, such as the Human Connectome Project (HCP) \cite{Tian2020}\cite{Karimi2022}. Diffusion data can vary widely due to different acquisition parameters, scanners, and patient populations and therefore unsupervised or self-supervised denoising methods are often preferred \cite{Fadnavis2020}. However, these self-supervised and unsupervised methods do not approach the performance of supervised techniques on data within the trained domain and can still perform variably with different out-of-domain datasets.

In addition, most dMRI denoising methods are evaluated only qualitatively or by denoising a subset of the data and evaluating the accuracy with respect to the full dataset. In this paper, we also evaluate denoising using external validation via test-retest reliability for Diffusion Tensor Imaging (DTI) \cite{Mukherjee2008} and Neurite Orientation Dispersion and Density Imaging (NODDI) \cite{Zhang2012} metrics to ensure precision as well as via known Structural Covariance Networks (SCNs) \cite{Wahl2010}\cite{Li2012} derived from those metrics in the white matter (WM) and gray matter (GM) to ensure biological accuracy. Finally, we pay special attention to the ability to remove heavy-tail noise, which can lead to biased biophysical metrics from fitting algorithms primarily designed for data corrupted with Gaussian noise.

Here we propose to use a Swin UNEt TRansformers (Swin UNETR) model \cite{Hatamizadeh2022} to denoise dMRI data conditioned on registered T1 scans. Unlike other supervised methods, which utilize a small subset of the HCP dataset (typically 40 subjects) for training \cite{Tian2020}\cite{Karimi2022}, we use the full set of HCP data, training with all b-values and diffusion-encoding directions, and apply simple data augmentations, such as random flipping, rotation, scaling, and k-space downsampling. Training on a large dataset, close to 300,000 3D volumes across 1021 subjects, allows the Swin UNETR model to learn a denoising function that generalizes well to many different conditions. In addition to the Swin UNETR, we also train a UNet convolutional neural network model that lacks a transformer component to determine the effect, if any, of network architecture on denoising performance. We validate our approach on a held-out HCP Retest dataset as well as three external datasets acquired in different patient populations using different scanners and dMRI protocols, showing significant benefits over current state-of-the-art self-supervised and unsupervised methods. In addition to improvements in accuracy, we also show better repeatability on the HCP Retest dataset. Finally, we demonstrate that fine-tuning, even on only one subject, improves performance on out-of-domain datasets and that our approach can also super-resolve dMRI data via qualitative assessment in an HCP subject.
\section{Methods}
\subsection{Data}
In our experiments, we used data from four different datasets the first of which comprising normal young adult volunteers is separated into a training dataset and a held-out dataset within this same training domain. The other three datasets are out-of-domain (OOD) patient research datasets for establishing generalizability and clinical applicability.
\begin{itemize}
  \item \textbf{HCP}: 1021 subjects from the Human Connectome Project (HCP) Young Adult dataset \cite{VanEssen2013} acquired using 90 diffusion-encoding directions at b-values of b=1000, 2000, 3000 s/mm\textsuperscript{2} with 1.25 mm resolution. We used all 1021 subjects for training and excluded any subjects in the HCP Retest dataset.
  \item \textbf{HCP Retest}: 44 subjects from the HCP Retest dataset used for validating denoising performance and repeatability. It was acquired in the same way as the HCP dataset.
  \item \textbf{TBI}: 45 adult mild traumatic brain injury patients acquired two decades ago using protocols identical to \cite{Wahl2010} on a 3T GE scanner with 55 diffusion-encoding directions at b=1000 s/mm\textsuperscript{2} with a nominal resolution of 1.8 mm (0.9 mm in xy after zero-interpolation in k-space). We randomly selected 5 subjects for fine-tuning and 40 subjects for testing.
  \item \textbf{SPIN}: 45 children ages 8-12 years with neurodevelopmental disorders acquired on a 3T Siemens Prisma scanner with 64 and 96 diffusion-encoding directions at b-values of \(b=1000, 2500\) s/mm\textsuperscript{2} respectively (TE=72.20 ms, TR=2420 ms, flip angle=85\(^{\circ}\)) with 2.00 mm isotropic resolution. We randomly selected 5 subjects for fine-tuning and 40 subjects for testing.
  \item \textbf{AHA}: 8 children and adolescents ages 8 - 18 with intracerebral hemorrhage due to vascular malformations, primarily arteriovenous malformations (AVMs), undergoing resection acquired on a 3T GE MR750 with 55 diffusion-encoding directions at b=2000 s/mm\textsuperscript{2} with a resolution of 2.00 mm that was zero-interpolated in-plane to 1.00 x 1.00 mm. Data was collected over three sessions: first, prior to resection, the second, six months after surgery, and the third, one year following surgery. Lesion location was determined by radiology notes. We selected a ninth subject for fine-tuning and had a total of 23 sessions for testing (one subject had only two sessions).
\end{itemize}
DMRI data was skull-stripped with Synthstrip \cite{Hoopes2022}, corrected for eddy current-induced distortions and subject movements with Eddy \cite{Andersson2016}, and aligned to structural 3D T1 scans with Boundary-Based Registration \cite{Greve2009}. T1 scans were also skull-stripped using Synthstrip and segmented using SynthSeg \cite{Billot2022}. For the AHA dataset, due to the hemorrhagic lesion impacting SynthSeg performance, T1 scans were instead segmented using Freesurfer's recon-all command with the SynthSeg option enabled and, in a few cases where recon-all failed, the recon-all-clinical command was used instead. Finally, co-registered T1 and dMRI scans were resampled with 5\textsuperscript{th} order spline interpolation at 1.25 mm and used as inputs to the model.
\subsection{Denoising Validation}
To evaluate the denoising algorithms and possible scan time speedup, we measure performance in both fully-sampled and subsampled data. We chose the minimum number of diffusion gradients necessary for unique fits: 6 for DTI, 15 for \nth{4} order spherical harmonic, and 28 for \nth{6} order spherical harmonic along with 1 b=0 s/mm\textsuperscript{2} volume. We select the directions to minimize the condition number of the design matrix using the procedure described in \cite{Tian2022}\cite{Skare2000}. For NODDI estimation, we only use the fully-sampled HCP acquisition. We compare the performance of the SWIN UNeTR and UNet models with three state-of-the-art unsupervised/self-supervised machine learning methods for dMRI denoising: block-matching and 4D filtering (BM4D) \cite{Maggioni2013}, Marchenko-Pastur Principal Component Analysis (MPPCA) \cite{Veraart2016}, and Patch2Self (P2S) \cite{Fadnavis2020}.

To evaluate DTI estimation, the mean absolute error (MAE) between the ground truth fully-sampled dataset and the model predictions in the subsampled dataset for the principal eigenvector (V1), fractional anisotropy (FA), axial diffusivity (AD), radial diffusivity (RD), and mean diffusivity (MD) were found. For evaluating higher order spherical harmonics, the Jensen–Shannon distance (JSD) between the ground truth and model predictions, projected onto a uniformly distributed 362 direction hemisphere, was used \cite{Cohen-Adad2011}. In each case, the ground truth was found by fitting the model using all acquired diffusion gradient directions. These errors were reported for WM and GM. DTI estimation was only conducted on the lowest shell (b=1000 s/mm\textsuperscript{2}) for multi-shell datasets, whereas spherical harmonic estimation was conducted on every shell. To evaluate super-resolution performance, dMRI data from an HCP subject was k-space downsampled by a factor of two and then upsampled to emulate a low resolution acquisition.

We explore denoising in an (OOD) dataset for clinical translation by applying our model on the AHA dataset, which consists of children with intracerebral hemorrhage undergoing lesion resection scanned before and after intervention. We collect global WM and GM DTI metrics and measure perilesional changes in DTI microstructure over time from both subsampled and fully-sampled shells. To investigate the output distribution and effect of signal rectification, we measure the signal in the lateral ventricles, a region which consists almost solely of free water and has a heavily-attenuated uniform signal, across all diffusion-encoding directions and compare the resulting signal intensity histograms produced by the denoising methods.

Microstructural repeatability was assessed using the HCP Retest dataset by measuring the within-subject coefficient of variation (CoV) for DTI and NODDI parameters across the two sessions. Briefly, we consider DTI test-retest reliability performance on both the fully-sampled and subsampled b=1000 s/mm\textsuperscript{2} shell. For NODDI estimation, we consider the full multi-shell HCP acquisition. To evaluate repeatability in WM regions, we conduct tract-based spatial statistics (TBSS) analysis via FA registration to a template using the Johns Hopkins University (JHU) atlas \cite{Smith2006}.

We also investigate the repeatability of SCNs by measuring the mean absolute difference between SCNs derived from the first and second sessions. We quantify both GM SCNs, which encompass only cortical GM regions from the Desikan-Killiany-Tourville atlas, and WM SCNs, which encompass only the JHU WM tracts given by TBSS analysis. We evaluate performance for DTI SCN repeatability on the subsampled b=1000 s/mm\textsuperscript{2} shell and for NODDI SCN repeatability on the fully-sampled acquisition using all shells. Finally, we also compute the MAE between the denoised subsampled DTI SCNs and the ground truth DTI SCN, computed by collecting the average DTI values for each region across both sessions using the fully-sampled b=1000 s/mm\textsuperscript{2} shell.

\subsection{Training and Implementation}
The Swin UNETR model \cite{Hatamizadeh2022}\cite{Liu2021} is implemented using PyTorch \cite{Paszke2019} and MONAI \cite{Cardoso2022} (Fig. \ref{fig:swin}). The model was trained on a NVIDIA V100 GPU using mean-squared error loss between the model output and ground truth. To obtain ground truth dMRI data for training, a \nth{6} order spherical harmonic was fit for each shell and projected onto the acquired directions. We chose AdamW \cite{Loshchilov2017} as an optimizer with a learning rate of 1e-5 and train using gradient clipping with a maximal norm of 1.0 and 16-bit precision for 14 epochs of training. During training, we first downsample the dMRI scan with a probability of 0.5 in frequency space to an anisotropic resolution between 1.25 and 3 mm and linearly upsample back to 1.25 mm resolution \cite{Lyu2019}. Random patches of 128 x 128 x 128 are cropped from the scan and randomly flipped with a probability of 0.5 along all axes and randomly rotated by 0, 90, 180, or 270 degrees along all axes with equal probability. The input dMRI patch is normalized to have zero mean and unit variance and the input T1 patch is normalized to have zero mean and standard deviation uniformly log-scaled between 0.25 and 4.0. For inference, we use a sliding window approach with an overlap of 0.875 and use \nth{5} order spline interpolation to upsample the data. Fine-tuning was achieved via additional training on external data from one held-out subject out of five with a learning rate of 1e-6 for three epochs and the average result for validation was reported. Finally, since our models are trained and evaluated on the HCP dataset with 1.25 mm resolution, model predictions are resampled to native dMRI resolution for external OOD dataset validation. UNet model training, fine-tuning, and validation was conducted in the same way, except training was extended to 20 epochs. Training for both the Swin and UNet models continued until training loss stabilized.
\section{Results}
\subsection{Denoising Performance}
\begin{table}
\centering
\caption{MAE of FA, MD, RD, AD, and V1 estimation using six-direction HCP, SPIN, TBI, and AHA data in white matter (WM) and gray matter (GM) via no denoising (RAW), P2S, MPPCA, BM4D, UNET, SWIN with no fine-tuning (SWIN), UNET with fine-tuning on one subject (UNET-F1), and SWIN with fine-tuning on one subject (SWIN-F1). Best results are \textbf{bolded}.}
\begin{tabular}{|c|c|c|c|c|c|c|c|c|c|c|}
\hline
Dataset & Tissue & Metric & RAW & P2S & MPPCA & BM4D & UNET & UNET-F1 & SWIN & SWIN-F1  
\\
\hline
\multirow{10}{*}{HCP} & \multirow{5}{*}{WM} & AD &    0.13 &   0.285 &            0.101 &           0.0848 &  0.0913 &  
& \textbf{0.0811} 
&
\\
    &    & FA &  0.0935 &   0.249 &           0.0672 &           0.0563 &  0.0556 &  
& \textbf{0.0496} 
&
\\
    &    & MD &  0.0657 &  0.0695 &           0.0556 &           0.0483 &  0.0526 &  
& \textbf{0.0452} 
&
\\
    &    & RD &  0.0773 &   0.124 &           0.0622 &           0.0541 &  0.0558 &  
& \textbf{0.0488} 
&
\\
    &    & V1 &    19.7 &    66.9 &             15.2 &             13.9 &    13.4 &    
& \textbf{12.6} 
&
\\
\cline{2-11}
    & \multirow{5}{*}{GM} & AD &   0.133 &   0.107 &           0.0903 &           0.0931 &  0.0957 &  
& \textbf{0.0794} 
&
\\
    &    & FA &   0.111 &  0.0793 &           0.0542 &           0.0588 &  0.0569 &  
& \textbf{0.0484} 
&
\\
    &    & MD &  0.0723 &  0.0728 &             0.07 &           0.0683 &  0.0725 &  
& \textbf{0.0597} 
&
\\
    &    & RD &  0.0849 &  0.0814 &           0.0731 &           0.0719 &  0.0754 &  
& \textbf{0.0626} 
&
\\
    &    & V1 &    33.6 &    64.1 &             28.7 &             28.1 &    28.4 &    
& \textbf{26.7} 
&
\\
\hline
\multirow{10}{*}{TBI} & \multirow{5}{*}{WM} & AD &   0.211 &   0.295 &            0.173 &            0.206 &   0.133 &            0.14 
& 0.131 
&\textbf{0.13}  
\\
    &    & FA &   0.145 &   0.229 &            0.117 &             0.14 &  0.0866 &           0.0864 
& 0.0922 
&\textbf{0.0808}  
\\
    &    & MD &  0.0941 &  0.0966 &           0.0886 &           0.0907 &  0.0743 &            0.0784 
& 0.073 
&\textbf{0.0702}  
\\
    &    & RD &   0.116 &   0.138 &            0.103 &            0.111 &  0.0841 &            0.0848 
& 0.085 
&\textbf{0.076}  
\\
    &    & V1 &    26.6 &    67.1 &             24.2 &             26.1 &    21.5 &             21.8 
& 22.1 
&\textbf{21.0}  
\\
\cline{2-11}
    & \multirow{5}{*}{GM} & AD &   0.218 &   0.136 &            0.173 &            0.217 &   0.147 &            0.151 
& 0.149 
&\textbf{0.127}  
\\
    &    & FA &   0.181 &  0.0966 &            0.139 &            0.175 &   0.102 &            0.102 
& 0.123 
&\textbf{0.0949}  
\\
    &    & MD &  0.0908 &  0.0897 &            0.089 &           0.0893 &  0.0905 &  0.0926 
& \textbf{0.0809} 
&0.0817  
\\
    &    & RD &   0.124 &  0.0982 &             0.11 &             0.12 &   0.102 &            0.104 
& 0.102 
&\textbf{0.0948}  
\\
    &    & V1 &    38.0 &    62.0 &             36.6 &             37.6 &    35.5 &             35.6 
& 36.2 
&\textbf{35.0}  
\\
\hline
\multirow{10}{*}{SPIN} & \multirow{5}{*}{WM} & AD &   0.112 &   0.594 &            0.099 &           0.0999 &  0.0869 &           0.0844 
& 0.0825 
&\textbf{0.0767}  
\\
    &    & FA &  0.0829 &   0.178 &           0.0768 &           0.0747 &  0.0609 &            0.0578 
& 0.057 
&\textbf{0.0518}  
\\
    &    & MD &  0.0553 &   0.203 &           0.0446 &           0.0438 &  0.0491 &            0.0415 
& 0.044 
&\textbf{0.0402}  
\\
    &    & RD &  0.0657 &   0.128 &           0.0561 &           0.0536 &   0.056 &           0.0479 
& 0.0504 
&\textbf{0.0457}  
\\
    &    & V1 &    18.4 &    63.9 &             16.9 &             15.0 &    14.8 &             14.6 
& 14.5 
&\textbf{13.8}  
\\
\cline{2-11}
    & \multirow{5}{*}{GM} & AD &   0.106 &   0.652 &           0.0759 &           0.0823 &  0.0905 &           0.0746 
& 0.0859 
&\textbf{0.0694}  
\\
    &    & FA &  0.0994 &   0.193 &           0.0571 &  \textbf{0.0508} &  0.0617 &           0.0559 
& 0.0636 
&0.0517  
\\
    &    & MD &  0.0527 &   0.207 &           0.0506 &           0.0652 &  0.0611 &            0.0494 
& 0.056 
&\textbf{0.0471}  
\\
    &    & RD &   0.066 &   0.106 &            0.057 &           0.0703 &  0.0642 &           0.0555 
& 0.0609 
&\textbf{0.0529}  
\\
    &    & V1 &    30.9 &    59.7 &             30.5 &             28.8 &    28.3 &             27.8 
& 28.3 
&\textbf{26.9}  
\\
\hline
\multirow{10}{*}{AHA} & \multirow{5}{*}{WM} & AD &   0.145 &   0.337 &           0.0981 &             0.14 &   0.091 &   0.0987 
& \textbf{0.088} 
&0.0921  
\\
    &    & FA &  0.0824 &   0.225 &           0.0661 &           0.0796 &    0.06 &           0.0736 
& 0.0607 
&\textbf{0.0591}  
\\
    &    & MD &  0.0407 &    0.15 &  \textbf{0.0312} &           0.0389 &  0.0378 &           0.0396 
& 0.0371 
&0.0362  
\\
    &    & RD &  0.0489 &   0.189 &           0.0378 &           0.0466 &   0.041 &           0.0502 
& 0.0414 
&\textbf{0.0374}  
\\
    &    & V1 &    20.1 &    69.2 &             18.1 &             19.4 &    16.6 &             16.8 
& 17.2 
&\textbf{16.0}  
\\
\cline{2-11}
    & \multirow{5}{*}{GM} & AD &   0.173 &   0.462 &            0.101 &            0.161 &  0.0866 &           0.0917 
& 0.0943 
&\textbf{0.0801}  
\\
    &    & FA &  0.0988 &   0.107 &           0.0685 &           0.0947 &  0.0569 &           0.0577 
& 0.0721 
&\textbf{0.0535}  
\\
    &    & MD &  0.0484 &   0.247 &  \textbf{0.0375} &           0.0449 &   0.047 &           0.057 
& 0.0445 
&0.0441  
\\
    &    & RD &  0.0611 &   0.183 &  \textbf{0.0454} &           0.0574 &  0.0497 &           0.0588 
& 0.0516 
&0.0464  
\\
    &    & V1 &    31.1 &    64.7 &             29.4 &             30.5 &    28.5 &             28.3 & 29.7 &\textbf{27.7}  \\
\hline
\end{tabular}
\label{table:dti_mae}
\end{table}

For DTI estimation, the Swin model achieves the lowest MAE in all metrics in WM and GM in the HCP and TBI validation datasets, even without any fine-tuning (Table \ref{table:dti_mae}). For the SPIN dataset, the Swin model also outperforms all other models, except for FA estimation in GM where BM4D achieved the best result. In the AHA dataset, with the exception of MD and RD, the Swin model achieves lower MAE than any other method. While the UNet performs better than BM4D, MPPCA, and P2S in most settings, its performance still lags behind the Swin model, especially the fine-tuned Swin model. Patch2Self performs worse than the other denoising methods in this setting, especially for V1 estimation. Qualitative comparisons are consistent with these quantitative results and show that the Swin model is able to capture more of the finer features in the WM and GM microstructure without the excessive smoothing of BM4D and MPPCA (Supplementary Figs. \ref{fig:hcp_denoise}, \ref{fig:spin_denoise}, \ref{fig:tbi_denoise}, \ref{fig:aha_denoise}).

\begin{figure}
    \centering
    \includegraphics[width=0.9\textwidth]{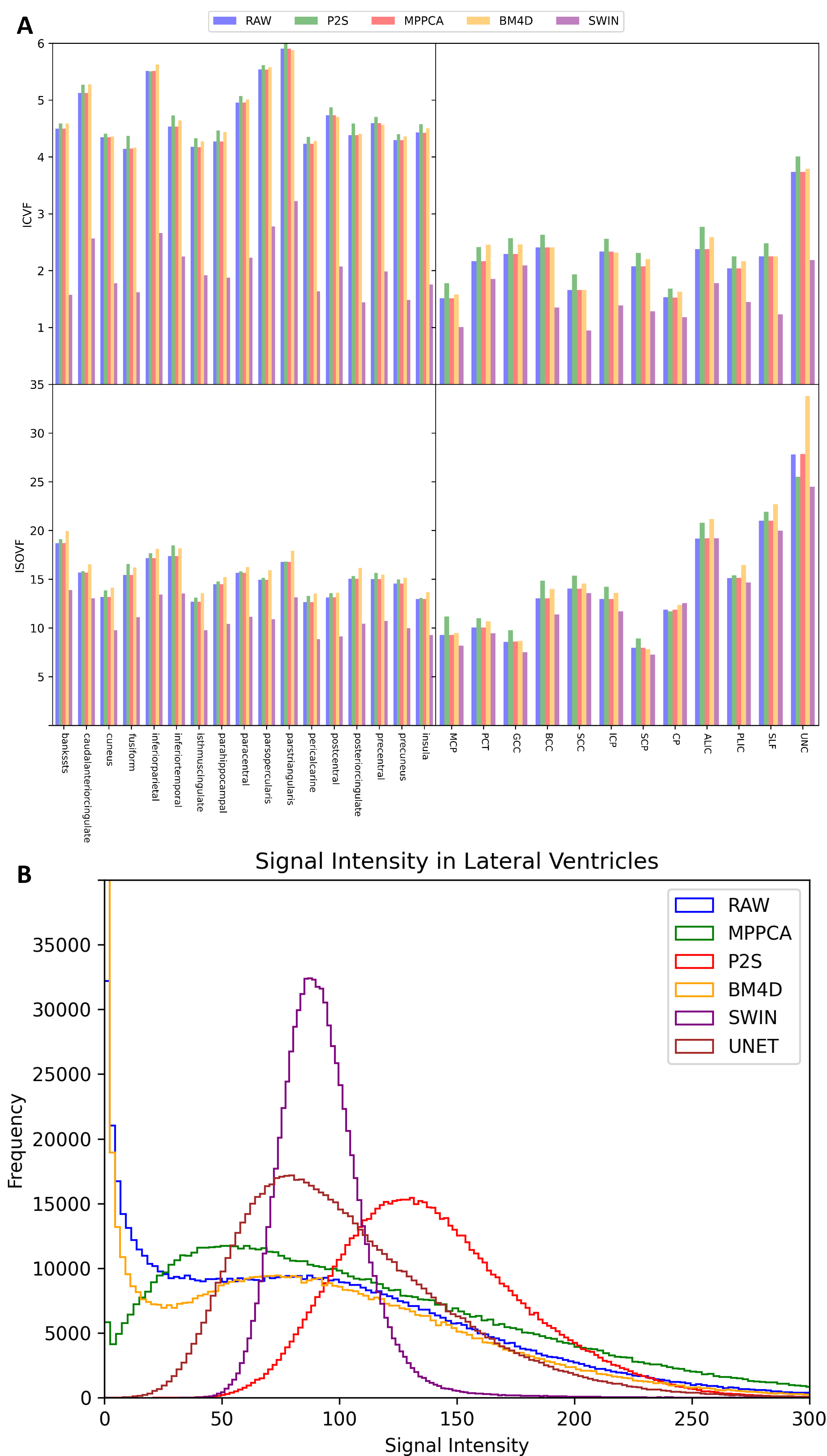}
    \caption{(A) ICVF and ISOVF CoV (\%) in select WM and GM regions and (B) histogram of signal intensity in the b=2000 s/mm\textsuperscript{2} shell in the lateral ventricles of an AHA patient. Results using no denoising (RAW), P2S, BM4D, MPPCA, SWIN, and UNET denoising are displayed.}
    \label{fig:hist_bar}
\end{figure}
Finally, we examine the noise distribution of each denoising algorithm in the lateral ventricles of one subject from the AHA dataset (Fig. \ref{fig:hist_bar}B). Unlike the other denoising algorithms, the Swin model is able to transform the original data distribution, a heavy-tailed Rician-like distribution, into a more Gaussian-like distribution with significantly smaller variance and skew but greater kurtosis.

\subsection{Test-Retest Reliability}
\begin{table}
\centering
\caption{Average CoV (\%) across all WM and GM regions for DTI estimation using 6 and 90 direction subsets of the B1000 shell and NODDI estimation using all available data. The MAE for DTI SCN estimation using 6 direction subsets and the repeatability error for DTI and NODDI SCNs using all available data. Results using no denoising (RAW), P2S, BM4D, MPPCA, and SWIN denoising are displayed. Best results are \textbf{bolded}.}
\begin{tabular}{|c|c|c|c|c|c|c|c|}
\hline
Validation & Tissue & Metric & RAW & P2S & MPPCA & BM4D & SWIN \\
\hline
\multirow{8}{*}{\makecell{6-Direction\\DTI\\CoV}} & \multirow{4}{*}{WM} & AD &           2.438 &            3.02 &           2.526 &   2.43 &  \textbf{2.339} \\
       &    & FA &  \textbf{2.603} &           7.492 &            3.09 &  2.912 &           2.922 \\
       &    & MD &           2.986 &           3.429 &           3.014 &  2.987 &  \textbf{2.901} \\
       &    & RD &           5.273 &  \textbf{4.009} &           5.136 &  5.123 &           4.846 \\
\cline{2-8}
       & \multirow{4}{*}{GM} & AD &           2.447 &           2.451 &            2.54 &  2.436 &  \textbf{2.295} \\
       &    & FA &           4.365 &           5.782 &           6.267 &    4.5 &  \textbf{4.186} \\
       &    & MD &           2.518 &           2.529 &           2.536 &  2.524 &  \textbf{2.436} \\
       &    & RD &           2.788 &           2.615 &            2.73 &  2.715 &  \textbf{2.611} \\
\hline
\multirow{8}{*}{\makecell{90-Direction\\DTI\\CoV}} & \multirow{4}{*}{WM} & AD &  \textbf{1.541} &           1.668 &           1.544 &  1.573 &           1.592 \\
       &    & FA &           2.484 &           2.527 &  \textbf{2.432} &  2.851 &           3.005 \\
       &    & MD &           1.573 &           1.702 &           1.577 &  1.562 &  \textbf{1.553} \\
       &    & RD &           1.649 &           1.772 &           1.653 &  1.617 &  \textbf{1.584} \\
\cline{2-8}
       & \multirow{4}{*}{GM} & AD &  \textbf{1.579} &           1.703 &           1.582 &  1.613 &           1.628 \\
       &    & FA &           2.614 &           2.651 &  \textbf{2.557} &  3.004 &           3.167 \\
       &    & MD &           1.624 &           1.748 &           1.627 &  1.613 &  \textbf{1.602} \\
       &    & RD &           1.694 &           1.809 &           1.698 &  1.659 &  \textbf{1.627} \\
\hline
\multirow{6}{*}{\makecell{NODDI\\CoV}} & \multirow{3}{*}{WM} & ICVF &  2.585 &          2.854 &  2.584 &  2.676 &   \textbf{1.68} \\
   &    & ODI &  5.349 &           6.93 &  5.389 &  5.617 &  \textbf{5.042} \\
   &    & ISOVF &  15.54 &          15.96 &  15.54 &  16.74 &  \textbf{14.41} \\
\cline{2-8}
& \multirow{3}{*}{GM} & ICVF &  4.943 &          5.068 &  4.943 &  5.012 &  \textbf{2.487} \\
   &    & ODI &  1.776 &  \textbf{1.77} &   1.78 &  1.982 &           1.792 \\
   &     & ISOVF &   14.9 &          15.28 &   14.9 &  15.65 &  \textbf{11.13} \\
\hline
\multirow{8}{*}{\makecell{6-Direction\\DTI\\SCN\\MAE}} & \multirow{4}{*}{WM} & AD &  0.3707 &           0.3186 &  \textbf{0.3164} &  0.3494 &           0.3273 \\
   & & FA &  0.2163 &           0.1815 &           0.1643 &  0.1879 &  \textbf{0.1544} \\
   & & MD &  0.2615 &           0.2691 &           0.2628 &  0.2632 &  \textbf{0.2478} \\
   & & RD &  0.1703 &           0.2034 &           0.1646 &  0.1689 &  \textbf{0.1593} \\
\cline{2-8}
& \multirow{4}{*}{GM} & AD &   0.182 &   \textbf{0.159} &           0.1654 &  0.1672 &           0.1613 \\
   & & FA &  0.2276 &           0.3128 &           0.2373 &  0.2104 &  \textbf{0.1763} \\
   & & MD &  0.1399 &  \textbf{0.1356} &           0.1381 &   0.138 &            0.136 \\
   & & RD &   0.127 &           0.1269 &           0.1306 &    0.13 &  \textbf{0.1265} \\
\hline
\multirow{12}{*}{\makecell{SCN\\Repeatability\\Error}} & \multirow{6}{*}{WM} & AD &  0.1413 &            0.1479 &  \textbf{0.1295} &  0.1384 &             0.134 \\
   & & FA &  0.1299 &            0.1231 &  \textbf{0.1203} &  0.1247 &            0.1207 \\
   & & MD &  0.1386 &            0.1428 &           0.1387 &  0.1415 &   \textbf{0.1378} \\
   & & RD &  0.1351 &             0.137 &  \textbf{0.1309} &  0.1354 &             0.133 \\
   & & ICVF &  0.2528 &            0.1891 &           0.2524 &  0.2759 &  \textbf{0.06437} \\
   & & ODI &  0.1364 &  \textbf{0.05801} &           0.1356 &  0.1388 &           0.07944 \\
   & & ISOVF &  0.3568 &            0.3329 &           0.3568 &  0.3689 &   \textbf{0.1298} \\
\cline{2-8}
& \multirow{6}{*}{GM} & AD &  0.1547 &   \textbf{0.1314} &           0.1737 &  0.1541 &            0.1494 \\
   & & FA &  0.1491 &   \textbf{0.1062} &           0.2125 &   0.178 &            0.1636 \\
   & & MD &  0.1296 &            0.1275 &           0.1293 &  0.1272 &   \textbf{0.1271} \\
   & & RD &  0.1253 &            0.1259 &           0.1239 &  0.1245 &   \textbf{0.1229} \\
   & & ICVF &  0.5988 &             0.565 &           0.5987 &  0.6192 &   \textbf{0.1017} \\
   & & ODI &   0.165 &            0.1688 &           0.1683 &  0.1652 &   \textbf{0.1142} \\
   & & ISOVF &  0.5177 &            0.4929 &           0.5177 &  0.5255 &   \textbf{0.1196} \\
\hline
\end{tabular}
\label{table:rep_summary}
\end{table}

The Swin model achieves the lowest CoV between the test and retest datasets for all metrics in GM and AD and MD in WM using the 6-direction subsampled shell as well as MD and RD in both GM and WM for the 90-direction fully-sampled shell (Table \ref{table:rep_summary}). When using 90 directions, applying no denoising leads to the lowest CoV for AD and MPPCA achieves the best result for FA estimation.

For repeatability of intracellular volume fraction (ICVF), fiber orientation dispersion index (ODI), and free water fraction (ISOVF), where all acquired data was used, Swin outperforms all other denoising methods except for ODI estimation in GM, where P2S is slightly better. In particular, Swin excels at ICVF repeatability, achieving close to 50 \% lower CoV than the next best method on average and achieves dramatically lower CoV on regional GM and WM measurements (Fig. \ref{fig:hist_bar}A). Swin also achieves considerably better test-retest reliability than the other denoising approaches for ISOVF as well, especially in global and regional GM measurements (Fig. 1A).

Swin generates the most accurate FA, MD, and RD SCN in WM and FA and RD SCN in GM. Patch2Self achieves the lowest error for AD and MD SCN in GM, while MPPCA has the most accurate AD SCN in WM. MPPCA achieves the lowest SCN repeatability error in WM for AD, FA, and RD estimation, while Patch2Self has the most repeatable SCN in WM for ODI estimation and GM for AD and FA estimation. Swin denoising has the lowest SCN repeatability error for MD, ICVF and ISOVF in WM as well as MD, RD, ICVF,  ODI and ISOVF in GM. Once again, Swin performed remarkably better than all competing methods for ICVF and ISOVF, achieving CoV values that are one-sixth to one-third of those from no denoising or denoising with P2S, MPPCA or BM4D.

\subsection{Clinical Validation}
\begin{figure}
    \centering
    \includegraphics[width=1.0\textwidth]{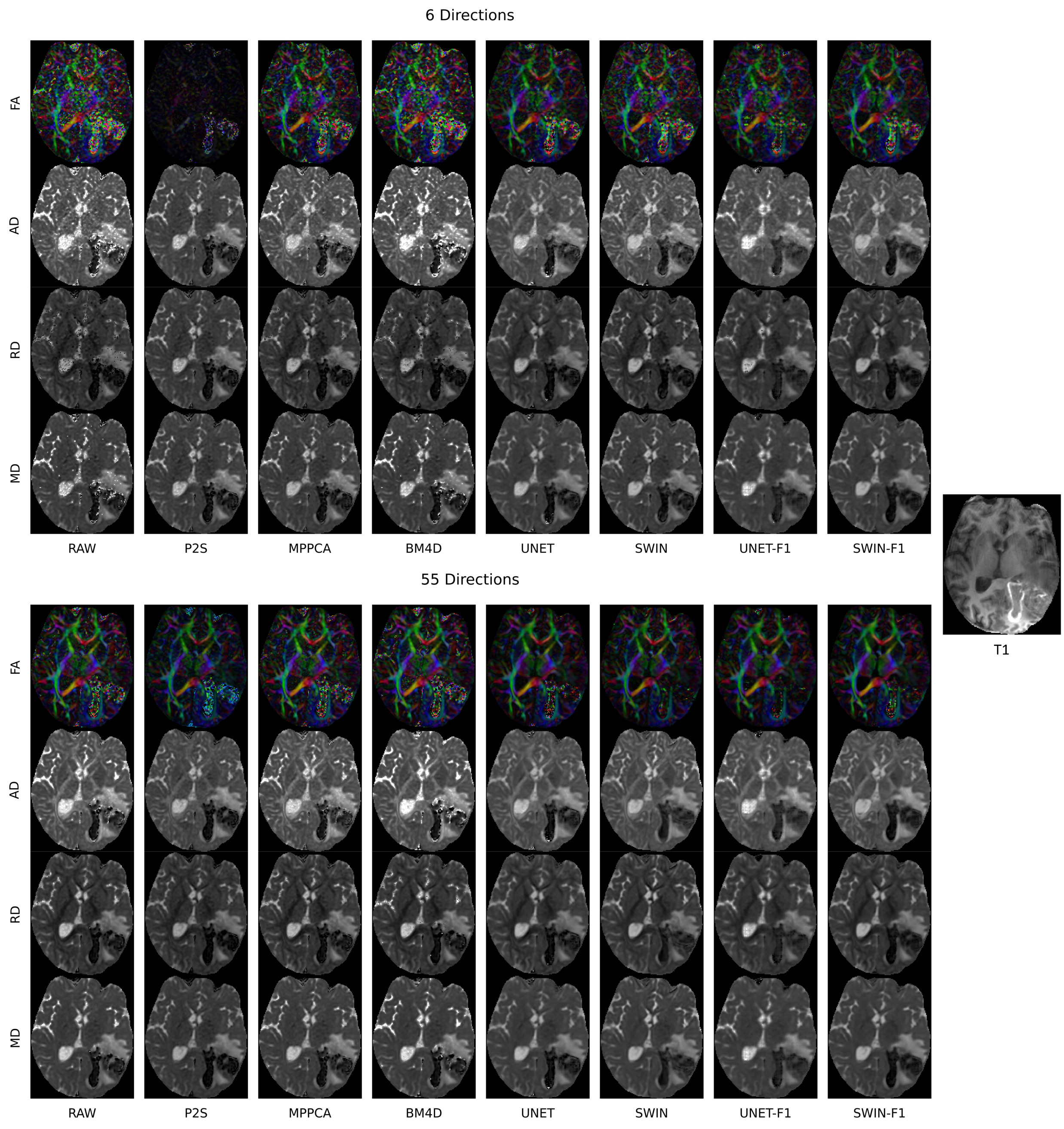}
    \caption{Effect of denoising on a 6 direction subset as well as a full shell (55 directions) for a session with poor quality data: contrast-to-noise ratio (CNR) = 1.2375 measured by Eddy in FSL. DTI metrics derived from no denoising (RAW), Patch2Self (P2S), MPPCA, BM4D, UNet denoising without fine-tuning (UNET), and Swin denoising without fine-tuning (SWIN-F1), UNet denoising with fine-tuning (UNET-F1), and Swin denoising with fine-tuning (SWIN-F1) are displayed as well as the T1 anatomical image.}
    \label{fig:aha_worst_case}
\end{figure}
The Swin model achieves qualitatively superior denoising even in poor quality data (CNR = 1.2375), which was the lowest CNR in the AHA dataset (Fig. \ref{fig:aha_worst_case}). Swin denoising conducted with only 6 directions approaches data quality achieved by acquiring all 55 directions, resulting in a 9-fold speedup of scan time. Applying Swin to all 55 directions removes much of the FA noise associated with the AVM and its surrounding hemorrhage. Fine-tuning leads to modest improvements in image quality for both the UNet and Swin models.

The Swin model, even with 6 directions, records DTI values which are consistent with those reported from other denoising algorithms with access to all 55 directions (Fig. \ref{fig:aha_boxplot}). Swin is able to achieve lower FA than the other methods with only 6 directions, barring P2S which yields unrealistically low values for WM, showing Swin's ability to reduce some of the noise which artifactually inflates FA. In addition, apart from FA estimation in GM, Swin denoising generates similar values for both 6-direction and 55-direction acquisitions, indicating that tissue microstructural metrics remain consistent, even as the angular resolution is increased. The Swin model also consistently produces lower FA, MD, RD, and AD values than other denoising methods in the perilesional space across all three sessions in all subjects (Fig. \ref{fig:aha_lesion_metrics}). MPPCA, BM4D, and no denoising (RAW) tend to follow the same values.

\subsection{Super-Resolution}
\begin{figure}
\centering
\includegraphics[width=1.0\textwidth]{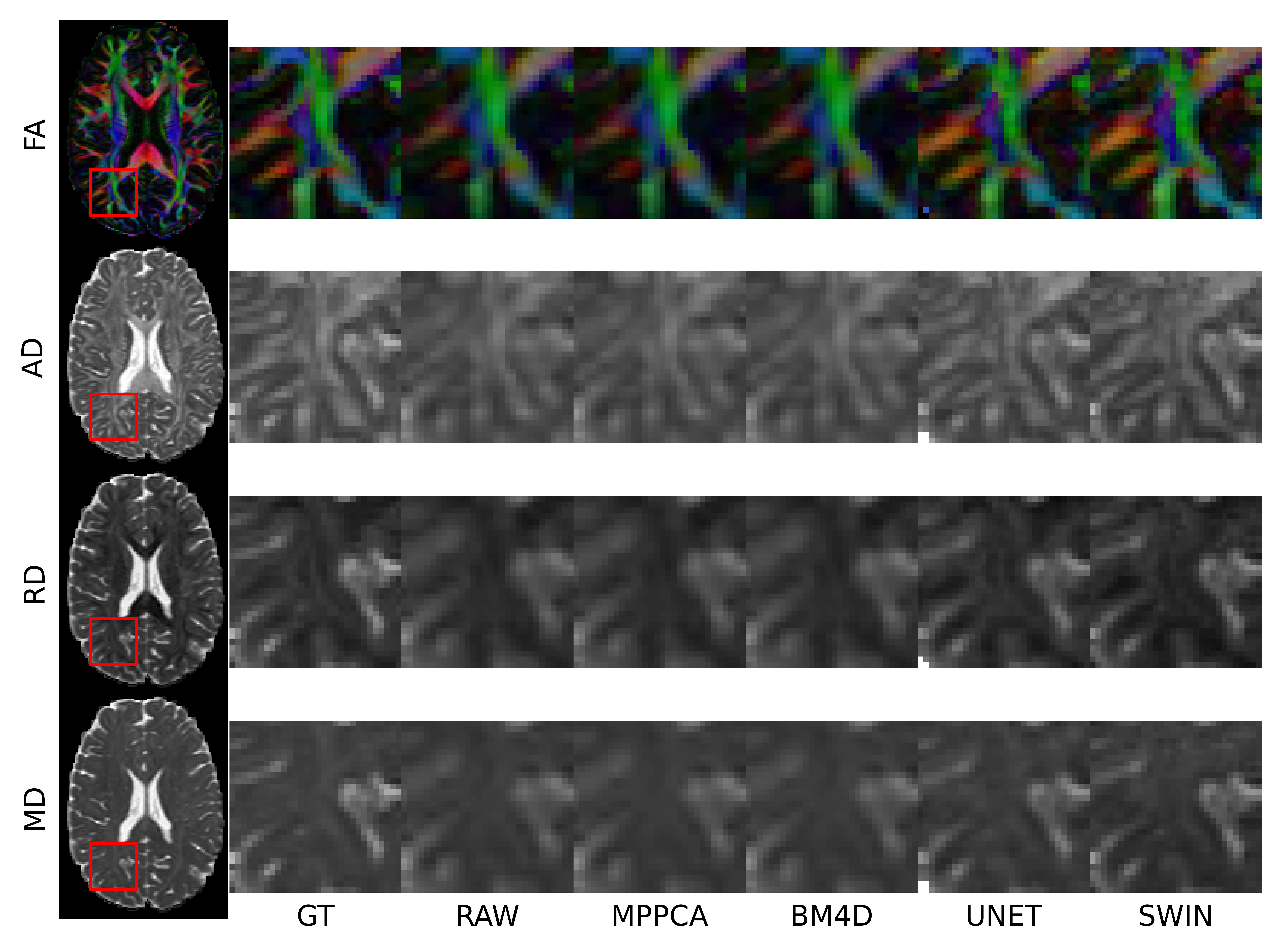}
\caption{Visual comparison between the ground truth (GT), no post-processing (RAW), MPPCA, BM4D, Unet, and Swin without fine-tuning (SWIN) for super-resolution in the posterior periventricular WM of an HCP subject. Data was k-space downsampled by a factor of two and then upsampled with \nth{5} order spline interpolation back to 1.25 mm.}
\label{fig:hcp_sr}
\end{figure}
Although our model was not trained for super-resolution, it can be used to resample a dMRI dataset to 1.25 mm resolution (Fig. \ref{fig:hcp_sr}). Qualitative comparison shows that Swin is able to capture more of the fine microstructure than BM4D and MPPCA in the posterior periventricular WM and avoids excessive blurring.

\section{Discussion}
To the best of the authors' knowledge, this is the first supervised dMRI denoising method that can be applied, without modification, to denoise dMRI datasets with widely varying scanners, patient populations, and acquisition parameters. We have validated our approach on a held-out portion of the HCP dataset as well as three external OOD datasets consisting of children with neurodevelopmental disorders, adults with TBI, and both children and adolescents with intracranial hemorrhage before and after resection of their vascular malformations. We have shown that our method can produce more accurate DTI metrics and spherical harmonic coefficients than other denoising methods with the minimum amount of data required for a unique fit. In addition, we have demonstrated the superior test-retest reliability of our denoising method for both DTI and NODDI metrics as well as SCNs derived from those metrics. Finally, we show that the Swin denoising has the potential to track brain microstructural changes with greater accuracy than other denoising models.

With Swin denoising, drastic scan time speedup is possible. Most dMRI protocols require at least 5 b=0 s/mm\textsuperscript{2} and 30 b=1000 s/mm\textsuperscript{2} acquisitions to obtain accurate DTI metrics, taking at least ten minutes \cite{Jones2004}. With Swin denoising, a dMRI acquisition which can obtain accurate DTI metrics is easily achievable in under two minutes. In fact, such a minimal 6-direction acquisition would take 90 seconds with an HCP protocol, 100 seconds with the TBI or AHA protocol, and only 20 seconds with the SPIN protocol. This speedup can allow dMRI to be used clinically in vulnerable populations and would significantly reduce motion artifacts that degrade image quality.

The Swin model showed high repeatability in all cases, but especially for the NODDI metrics of tissue intracellular volume fraction and free water fraction, often achieving better than 50\% lower CoV than the next best method. We believe that this could be due to the ability of the Swin model to have an approximately Gaussian output distribution. Unlike other denoising methods, the Swin model is able to remove the heavy tail noise inherent in dMRI data. Even BM4D, which is designed to correct for Rician noise, fails at this task. We believe that the Swin model and to a lesser extent the Unet succeed in this respect partly because, unlike other denoising models, they have access to the T1 anatomical image volume which aids in tissue segmentation, differentiating GM from WM from CSF. In addition, both models are trained to reduce mean-squared error, which heavily penalizes outliers and thus reduces the heavy tail.  

There were cases where the Swin model was not the best. For instance, the Swin model performs worse than MPPCA in WM for \nth{6} order spherical harmonic fitting, even in HCP data, possibly because the Swin model denoises one direction at a time, whereas MPPCA is able to collectively denoise all directions (Table \ref{table:sphm_jsd}). By processing each dMRI volume separately, we are able to consider the full brain volume and avoid artifact propagation across volumes but do not utilize the correlation across volumes, such as the transformer patch-based approach in \cite{Karimi2022}. Due to GPU memory constraints, a trade-off exists between utilizing spatial and angular correlations. Data compression techniques, such as quantized variational auto-encoders, should be explored to overcome this obstacle \cite{Duan2023}.

Our results were achieved with simple data augmentations, although using more extensive simulations, such as those in \cite{Muckley2021a}, could lead to greater generalizability. In addition, all inputs had to be resampled to 1.25 mm isotropic resolution, but with further data augmentation or resolution-independent architectures \cite{Dehghani2023}, it may be possible to perform denoising on native resolution if super-resolution is not desired. Our results require a T1 anatomical scan to be acquired in conjunction with the dMRI data. While most datasets which have dMRI data also collect T1 scans, this might not always the case and further work should be conducted to determine how important adding anatomical information, in the form of a T1 scan, is to denoising performance and whether T2 and other sequences can provide complementary information and further improve performance. Finally, our denoising method is designed to be performed after correction for patient movement, susceptibility-induced distortions, and eddy-current correction via the Eddy command as well as co-registration with a T1 scan. However, it may be fruitful to pursue 2D denoising which can clean raw data slice-by-slice to improve the performance of Eddy itself and yield better downstream results.

In addition, we trained a simple UNet which, while performing better than the other self-supervised methods, did not outperform the Swin model. This could be due to the ability of the Swin transformers to capture long-range dependencies better than a UNet, especially when provided with enough data \cite{Dosovitskiy2020}. It could also be because the same hyper-parameters were used for both the Swin and the UNet and those hyper-parameters were simply more optimal for the Swin model. Further investigation, with extensive hyper-parameter tuning on various datasets, would be necessary to determine the optimal architecture. In principle, we believe that most advanced neural network architectures, with sufficient complexity, would perform adequately.

We believe that some of the generalizability of our model may be attributed to grokking, a phenomenon that occurs when a neural network with good training loss but poor generalisation will, upon further training, transition to perfect generalisation \cite{Power2022}. We anecdotally observed grokking to occur after the sixth epoch of training and suspect this is due to using weight decay and AdamW adaptive stochastic gradient descent training on a sufficiently large dataset with data augmentation. A better understanding of what leads to grokking could be instrumental in designing generative AI models that generalize well at scale for healthcare.

Finally, fine-tuning on even one subject consistently led to improved denoising performance for DTI fitting, but the results were mixed in higher order spherical harmonic fitting. This could be because fine-tuning reduces model bias, but increases variance which can accumulate error over the 15 or 28 directions used to compute the 4th or 6th order spherical harmonics, respectively, compared to the six directions used in DTI estimation. In addition, we did not experience significant performance benefits by fine-tuning on more than one subject. Recent work has demonstrated that large language models can effectively learn from a single example and our model, while significantly smaller, might also exhibit similar behavior via successful domain adaptation with limited fine-tuning data. To the best of our knowledge, fine-tuning has never been reported before in the context of dMRI denoising and merits further investigation.

\textbf{Acknowledgements}. HCP data were provided by the Human Connectome Project, WU-Minn Consortium (Principal Investigators: David Van Essen and Kamil Ugurbil; U54 MH091657) funded by the 16 NIH Institutes and Centers that support the NIH Blueprint for Neuroscience Research; and by the McDonnell Center for Systems Neuroscience at Washington University. TBI data were acquired as part of a research project funded by NIH R01NS060886 (Principal Investigator: Pratik Mukherjee). SPIN data were acquired as part of a research project funded by NIH R01 MH116950 (Principal Investigators: Pratik Mukherjee and Elysa J. Marco). AHA data were acquired with funding from the American Heart Association (AHA) Bugher Foundation (Principal Investigators: Heather Fullerton, Christine Fox, Helen Kim, and Pratik Mukherjee).

%
%
%
%
\bibliographystyle{splncs04}
\bibliography{paper.bib}
\end{document}


\title{Generative AI for Rapid Diffusion MRI with Improved Image Quality, Reliability and Generalizability Supplementary}
\titlerunning{Generative AI for Rapid dMRI Supplementary}
\author{Amir Sadikov\inst{1}\inst{2} \and Xinlei Pan\inst{3} \and Hannah Choi\inst{1} \and Lanya T. Cai\inst{1} \and Pratik Mukherjee\inst{1}\inst{2}}
\authorrunning{Sadikov et al.}
%
\institute{Radiology and Biomedical Imaging, University of California, San Francisco \and Graduate Group in Bioengineering, University of California, San Francisco \and University of California, Berkeley\\
\email{amir.sadikov@ucsf.edu}}
\maketitle              
\section{Introduction}
For a list of all abbreviations used in the manuscript and their definitions, see Table \ref{table:acronym}.
\begin{longtable}{| p{.3\textwidth} | p{.7\textwidth} |}
\caption{Table of Abbreviations}\label{table:acronym}\\
\hline
Abbreviation & Definition \\
\hline
dMRI & Diffusion MRI\\
DTI & Diffusion Tensor Imaging\\
NODDI & Neurite Orientation Dispersion and Density Imaging\\
SCN & Structural Covariance Network\\
HCP & Human Connectome Project\\
AHA & American Heart Association\\
TBSS & Tract-Based Spatial Statistics\\
JHU & Johns Hopkins University\\
MAE & Mean Absolute Error\\
JSD & Jensen–Shannon Distance\\
CoV &  Coefficient of Variation\\
SNR &  Signal-to-Noise Ratio\\
CNR & Contrast-to-Noise Ratio\\
OOD &  Out-Of-Domain\\
WM &  White Matter\\
GM & Gray Matter\\
GT & Ground Truth\\
RAW & No Denoising Applied\\
BM4D & Block-Matching and 4D filtering\\
MPPCA & Marchenko-Pastur Principal Component Analysis\\
P2S & Patch2Self\\
UNET-F1 & UNET with fine-tuning on one subject\\
Swin UNETR/SWIN & Swin UNEt TRansformers\\
SWIN-F1 & SWIN with fine-tuning on one subject\\
V1 & Principal Eigenvector\\
FA &  Fractional Anisotropy\\
AD & Axial Diffusivity\\
RD & Radial Diffusivity\\
MD & Mean Diffusivity \\
ICVF & Intracellular Volume Fraction\\
ODI & fiber Orientation Dispersion Index\\
ISOVF & Free Water Fraction\\
MCP & Middle Cerebellar Peduncle\\
PCT & Pontine Crossing Tract\\
GCC & Genu of Corpus Callosum\\
BCC & Body of Corpus callosum\\
SCC & Splenium of corpus callosum\\
FX & Fornix\\
CST & Corticospinal Tract\\
ML & Medial Lemniscus\\
ICP & Inferior Cerebellar Peduncle\\
SCP & Superior Cerebellar Peduncle\\
CP & Cerebral Peduncle\\
ALIC & Anterior Limb of Internal Capsule\\
PLIC & Posterior Limb of Internal Capsule\\
RLIC & Retrolenticular Limb of Internal Capsule\\
ACR & Anterior Corona Radiata\\
SCR & Superior Corona Radiata\\
PCR & Posterior Corona Radiata\\
PTR & Posterior Thalamic Radiation\\
SS & Sagittal Stratum\\
EC & External Capsule\\
CGC & Cingulum (Cingulate Gyrus)\\
CGH & Cingulum (Parahippocampal)\\
FXST & fornix and Stria Terminalis\\
SLF & Superior Longitudinal Fasciculus\\
SFO & Superior Fronto-Occipital Fasciculus\\
UNC & Uncinate Fasciculus\\
TPT & Tapetum\\
\hline
\end{longtable}

\section{Methods}
A mask for the perilesional space was found by taking taking the largest connected component of the region that Freesurfer \textit{recon-all} segments as \textit{"unknown"} inside the brain mask. We dilate this mask two-fold and only take the dilated portion to be perilesional.
\section{Results}
\begin{figure}
    \centering
    \includegraphics[width=0.9\textwidth]{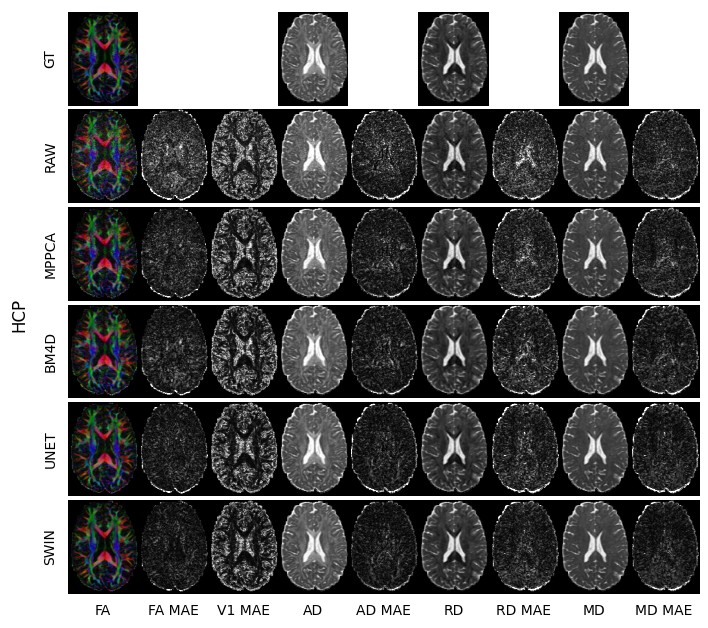}
    \caption{Visual comparison between the ground truth (GT), no denoising (RAW), BM4D, MPPCA, Unet, and Swin without finetuning (SWIN) for denoising on validation data from the HCP dataset}
    \label{fig:hcp_denoise}
\end{figure}
\begin{figure}
    \centering
    \includegraphics[width=0.9\textwidth]{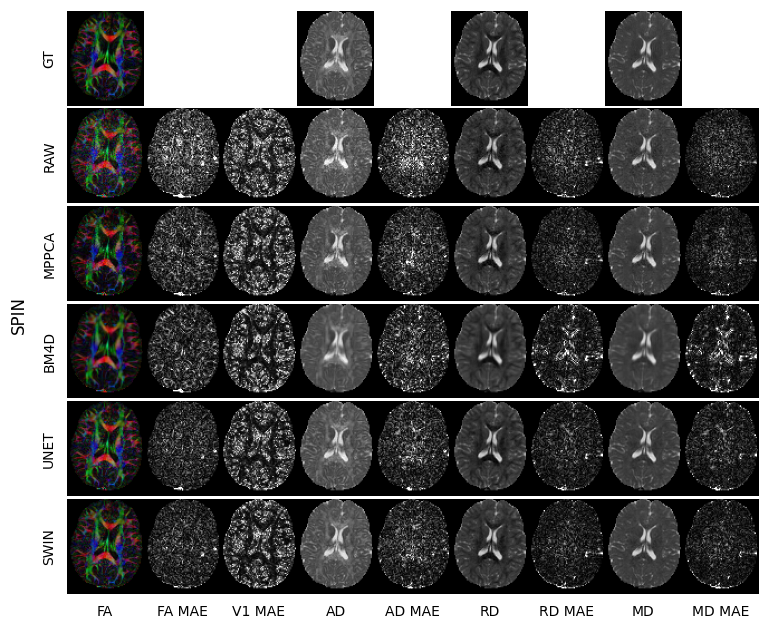}
    \caption{Visual comparison between the ground truth (GT), no denoising (RAW), BM4D, MPPCA, Unet, and Swin without finetuning (SWIN) for denoising on validation data from the SPIN dataset}
    \label{fig:spin_denoise}
\end{figure}
\begin{figure}
    \centering
    \includegraphics[width=0.9\textwidth]{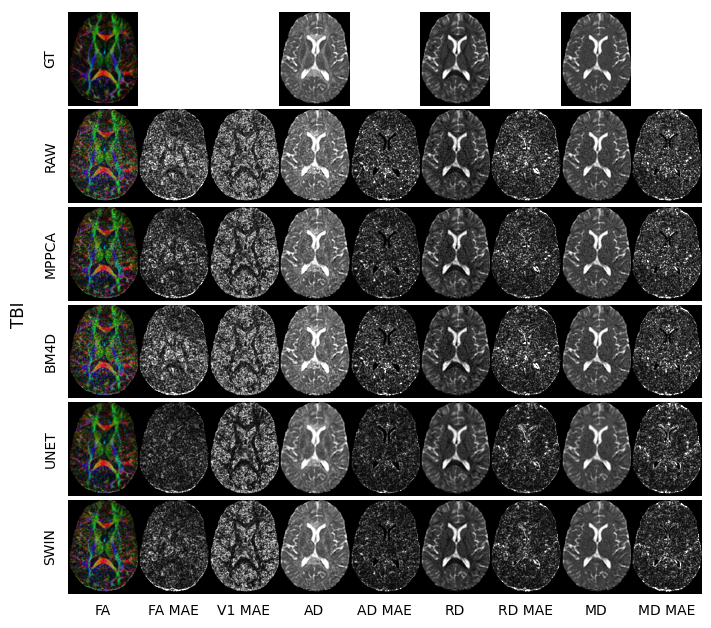}
    \caption{Visual comparison between the ground truth (GT), no denoising (RAW), BM4D, MPPCA, UNET, and Swin without finetuning (SWIN) for denoising on validation data from the TBI dataset}
    \label{fig:tbi_denoise}
\end{figure}
\begin{figure}
    \centering
    \includegraphics[width=0.9\textwidth]{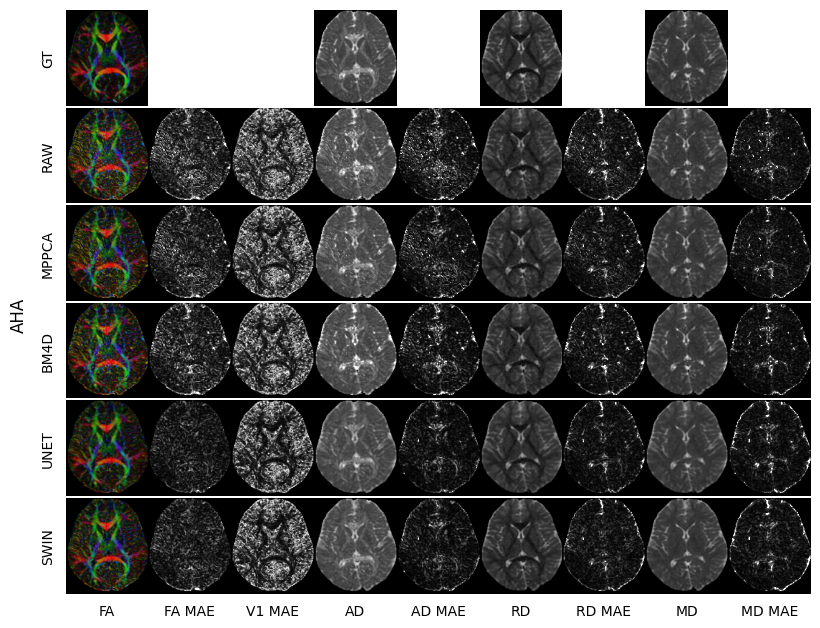}
    \caption{Visual comparison between the ground truth (GT), no denoising (RAW), BM4D, MPPCA, UNET, and Swin without finetuning (SWIN) for denoising on validation data from the AHA dataset}
    \label{fig:aha_denoise}
\end{figure}
\begin{figure}
\centering
\includegraphics[width=0.9\textwidth]{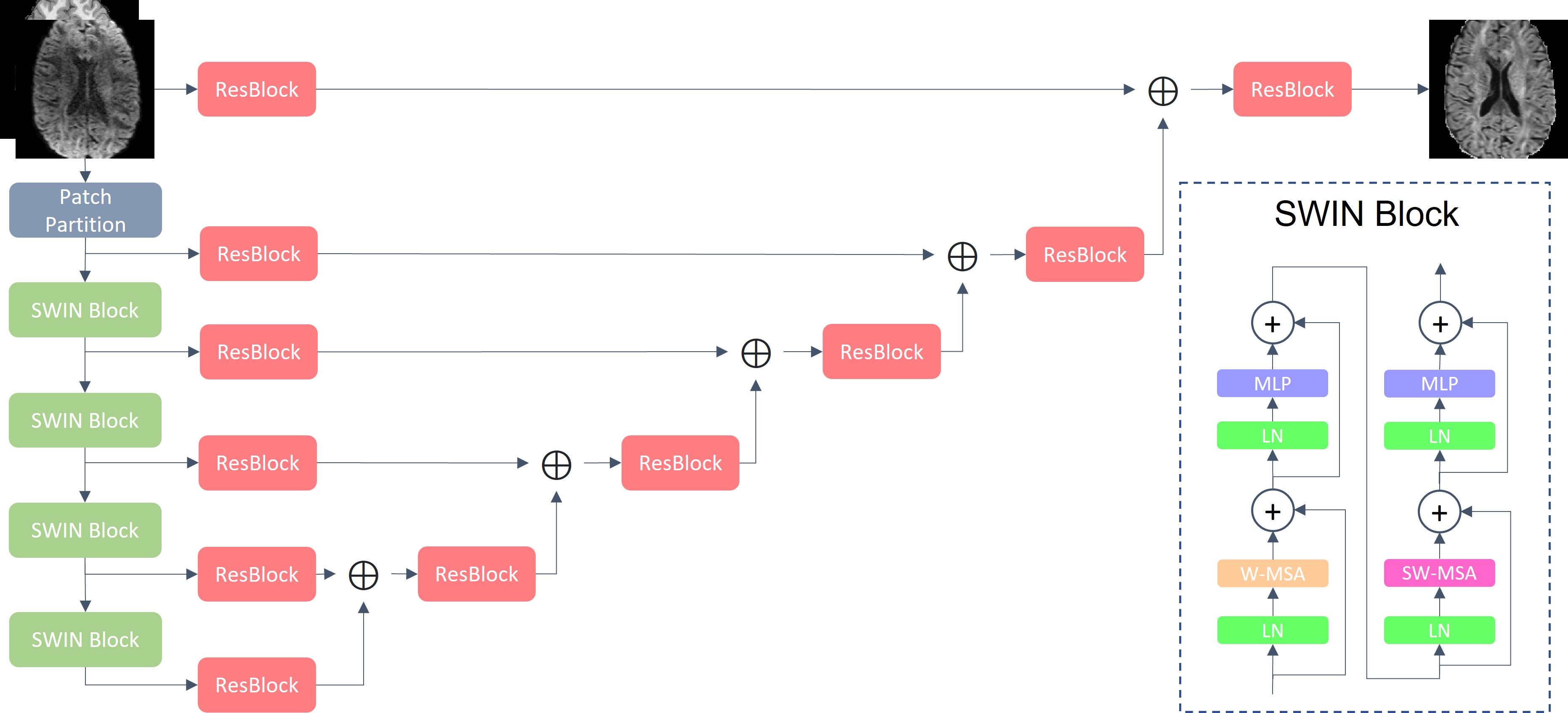}
\caption{A general overview of the Swin-UNETR architecture. The input is a concatenated 3D T1 and dMRI scan, which is encoded by a Swin transformer at multiple resolutions and fed into a residual convolutional neural net (CNN) decoder to reconstruct the ground truth dMRI scan.}
\label{fig:swin}
\end{figure}
\begin{figure}
    \centering
    \includegraphics[width=1.0\textwidth]{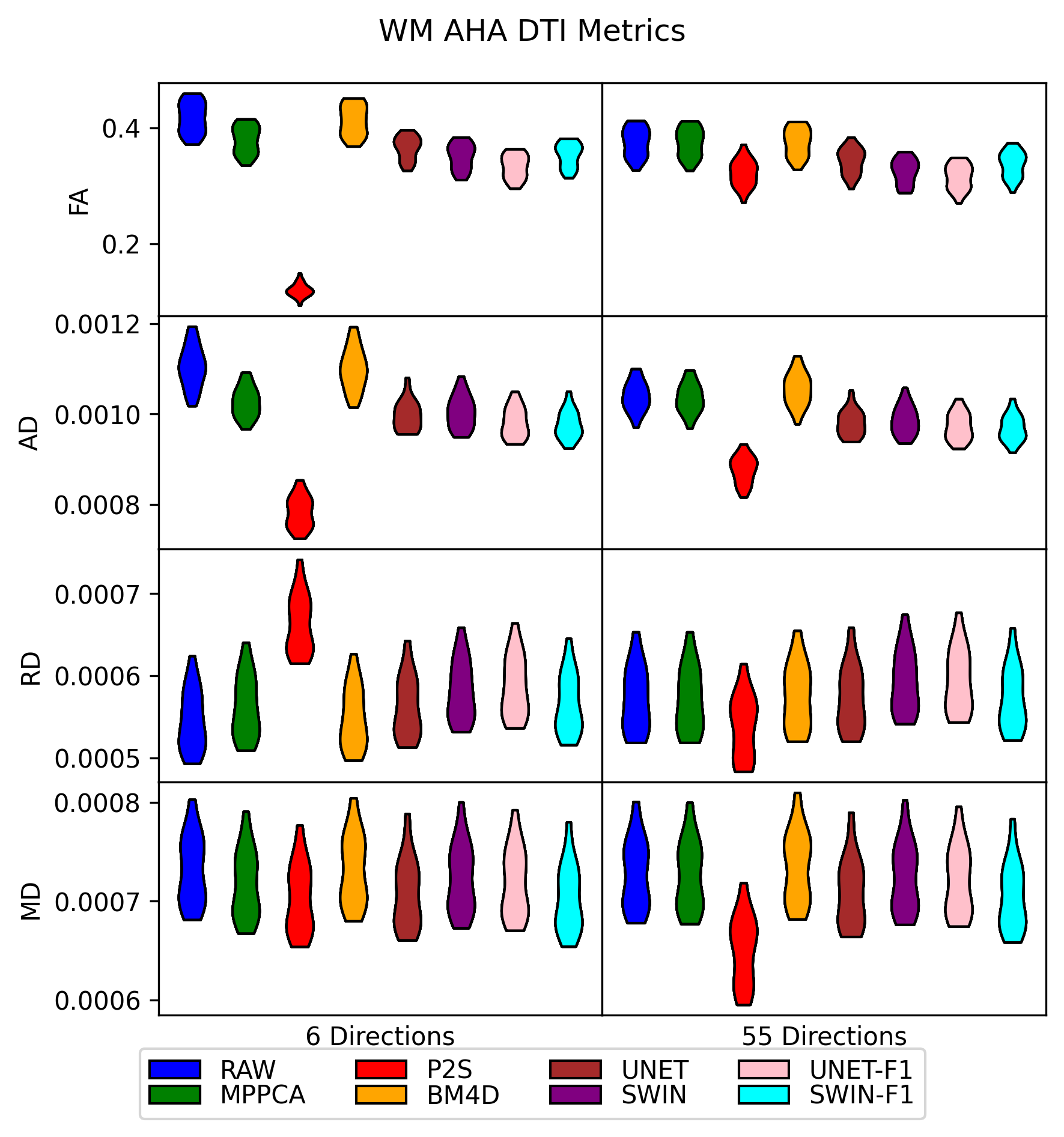}
    \caption{DTI metrics in WM and GM from no denoising (RAW), MPPCA, BM4D, and Swin denoising for 6 direction and 55 direction subsets from the AHA b=2000 s/mm\textsuperscript{2} shell.}
    \label{fig:aha_boxplot}
\end{figure}
\begin{figure}
    \centering
    \includegraphics[width=0.8\textwidth]{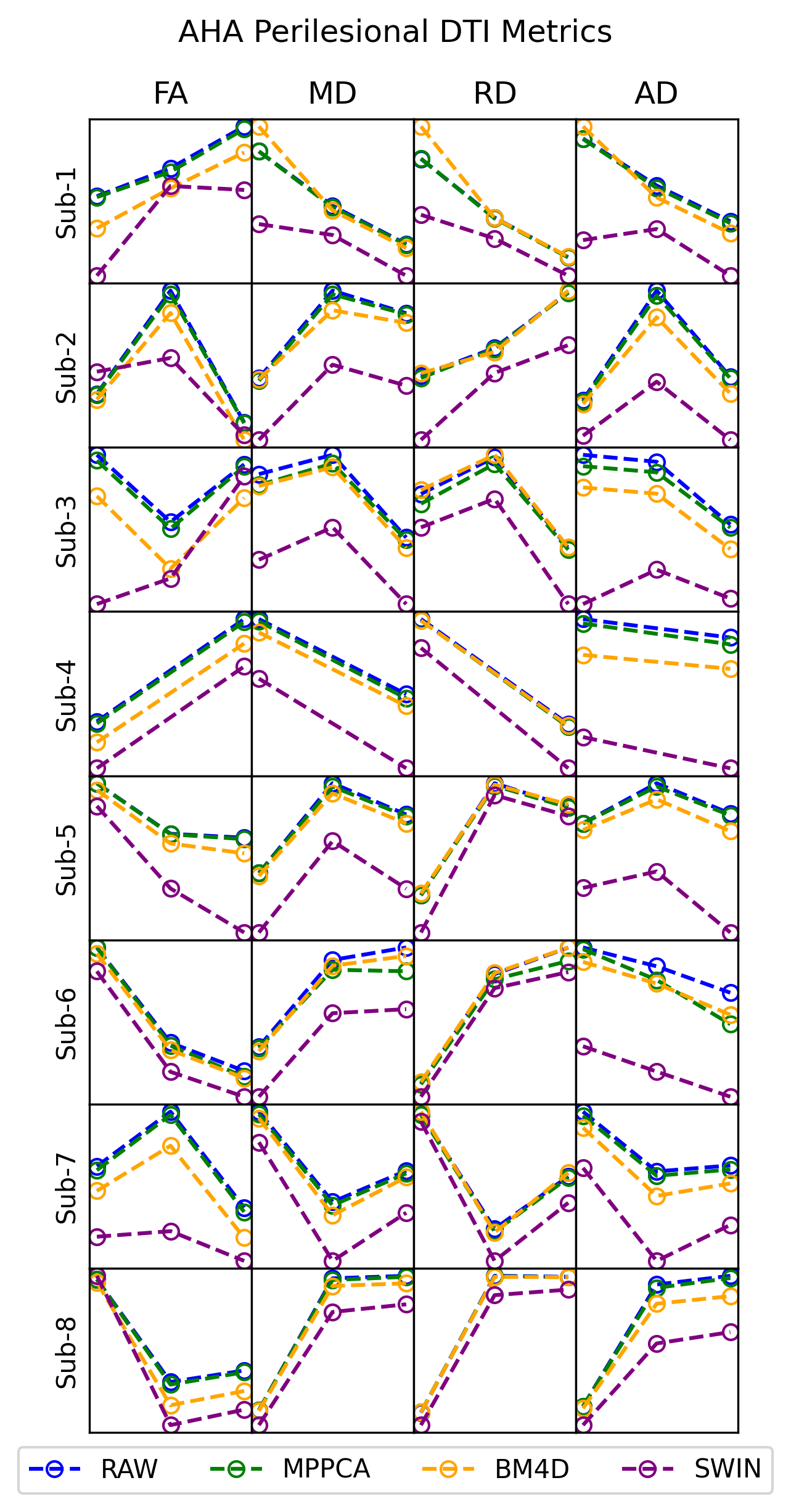}
    \caption{Average DTI metrics (AD, FA, MD, RD) from denoising dMRI data in the perilesional space of AHA subjects across three sessions: first session data is taken one day prior to AVM resection, second session data is taken 6 months after AVM resection, and third session is taken one year after AVM resection. Subject 4 has data from only sessions one and three.}
    \label{fig:aha_lesion_metrics}
\end{figure}

For \nth{4} and \nth{6} order spherical harmonic fitting, the Swin model achieves the lowest JSD in GM across all datasets (Table \ref{table:sphm_jsd}). In particular, the Swin model outperforms all other denoising methods in the external datasets (AHA, TBI, SPIN) with the exception of 6th order estimation of B1000 WM spherical harmonics in SPIN and B2000 WM spherical harmonics in AHA, where no denoising has the lowest JSD. For the HCP dataset, the MPPCA had a lower JSD than the Swin model for 6th order spherical harmonic estimation in higher shells. The Unet model achieves lower JSD than the Swin model in several metrics in the HCP dataset, but performs worse in the external datasets.
\begin{table}
\centering
\caption{JSD between ground truth and estimation using 15-direction (4th order) and 28-direction (6th order) HCP, SPIN, TBI, and AHA data in white matter (WM) and gray matter (GM) via no denoising (RAW), P2S, BM4D, MPPCA, UNET, and SWIN (with and without no finetuning). Best results are \textbf{bolded}.}
\begin{tabular}{|c|c|c|c|c|c|c|c|c|c|c|c|}
\hline
Site & Order & Shell & Tissue & RAW & P2S & MPPCA & BM4D & UNET & SWIN & UNET-F1 & SWIN-F1 \\
\hline
\multirow{12}{*}{HCP} & \multirow{6}{*}{4th} & \multirow{2}{*}{B1000} & WM &           0.0279 &  0.0385 &          0.0264 &  0.0257 &  \textbf{0.0246} &           0.0251 &         &                  \\
    &     &       & GM &           0.0232 &   0.023 &          0.0208 &  0.0208 &            0.019 &  \textbf{0.0184} &         &                  \\
\cline{3-12}
    &     & \multirow{2}{*}{B2000} & WM &           0.0423 &  0.0627 &          0.0396 &  0.0382 &   \textbf{0.038} &           0.0383 &         &                  \\
    &     &       & GM &           0.0365 &  0.0396 &          0.0317 &  0.0309 &           0.0292 &  \textbf{0.0286} &         &                  \\
\cline{3-12}
    &     & \multirow{2}{*}{B3000} & WM &           0.0522 &  0.0742 &          0.0477 &  0.0459 &           0.0461 &  \textbf{0.0458} &         &                  \\
    &     &       & GM &           0.0486 &   0.047 &          0.0401 &   0.039 &           0.0371 &  \textbf{0.0364} &         &                  \\
\cline{2-12}
    & \multirow{6}{*}{6th} & \multirow{2}{*}{B1000} & WM &           0.0206 &  0.0268 &          0.0203 &  0.0209 &  \textbf{0.0199} &           0.0205 &         &                  \\
\hline
\end{tabular}
\label{table:sphm_jsd}
\end{table}

The Swin model is able to achieve lower byte-lengths of dMRI data compressed with zlib using both high and low compression levels (Table \ref{table:file_size}). 
\begin{table}
\centering
\caption{The sum of all dMRI data byte-lengths in Gb for the HCP test-retest diffusion dataset undergoing no denoising (RAW), Patch2Self (P2S), MPPCA, BM4D, and Swin denoising compressed with lzib, using the DEFLATE algorithm, with compression levels of 1 (lowest) and 9 (highest). Best results are \textbf{bolded}.}
\begin{tabular}{|c|c|c|c|c|c|}
\hline
Compression Level & RAW & P2S & MPPCA & BM4D & SWIN \\
\hline
1 & 64.3381 & 64.0960 & 64.3384 & 64.1420 & \textbf{64.0651} \\
\hline
9 & 62.5757 & 62.2917 & 62.5755 & 62.3479 & \textbf{62.2508} \\
\hline
\end{tabular}
\label{table:file_size}
\end{table}
\begin{longtable}{| p{.34\textwidth} | p{.11\textwidth} | p{.11\textwidth} | p{.11\textwidth} | p{.11\textwidth} | p{.11\textwidth} | p{.11\textwidth} |}
\caption{Coefficient of variation (\%) for FA estimation via no denoising (RAW), P2S, BM4D, MPPCA, and SWIN for gray matter cortical regions in HCP test-retest data. Best results are \textbf{bolded}.}\label{table:repeat_fa_gm}\\
\hline
Cortical Regions & GT & RAW & P2S & MPPCA &  BM4D & SWIN \\
\hline
lh-bankssts                 &  2.172 &           4.164 &  5.232 &  5.394 &           4.023 &   \textbf{3.46} \\
lh-caudalanteriorcingulate  &  2.802 &           5.007 &  6.701 &  6.667 &           4.382 &  \textbf{4.095} \\
lh-caudalmiddlefrontal      &  3.514 &           4.353 &  5.843 &  5.293 &           3.614 &  \textbf{3.521} \\
lh-cuneus                   &  2.675 &           4.051 &   5.03 &  6.597 &            4.95 &  \textbf{3.906} \\
lh-entorhinal               &  3.157 &  \textbf{4.179} &  7.537 &   8.49 &           5.477 &           5.085 \\
lh-fusiform                 &  1.889 &           3.047 &  4.621 &  5.696 &           3.455 &  \textbf{3.032} \\
\hline
\end{longtable}
\begin{longtable}{| p{.34\textwidth} | p{.11\textwidth} | p{.11\textwidth} | p{.11\textwidth} | p{.11\textwidth} | p{.11\textwidth} | p{.11\textwidth} |}
\caption{Coefficient of variation (\%) for MD estimation via no denoising (RAW), P2S, BM4D, MPPCA, and SWIN for gray matter cortical regions in HCP test-retest data. Best results are \textbf{bolded}.}\label{table:repeat_md_gm}\\
\hline
Cortical Regions & GT & RAW & P2S & MPPCA &  BM4D & SWIN \\
\hline
lh-bankssts                 &   1.072 &           2.048 &            2.02 &  2.054 &           2.022 &  \textbf{1.957} \\
lh-caudalanteriorcingulate  &   1.677 &           2.419 &           2.409 &  2.427 &           2.361 &  \textbf{2.331} \\
lh-caudalmiddlefrontal      &   1.333 &           2.277 &           2.308 &  2.274 &           2.243 &  \textbf{2.122} \\
lh-cuneus                   &   1.498 &           2.096 &   \textbf{2.08} &  2.092 &           2.081 &           2.108 \\
lh-entorhinal               &   2.644 &  \textbf{4.279} &           4.369 &  4.381 &           4.351 &           4.296 \\
lh-fusiform                 &  0.9479 &           1.732 &           1.759 &  1.755 &           1.776 &   \textbf{1.67} \\
\hline
\end{longtable}
\begin{longtable}{| p{.34\textwidth} | p{.11\textwidth} | p{.11\textwidth} | p{.11\textwidth} | p{.11\textwidth} | p{.11\textwidth} | p{.11\textwidth} |}
\caption{Coefficient of variation (\%) for RD estimation via no denoising (RAW), P2S, BM4D, MPPCA, and SWIN for gray matter cortical regions in HCP test-retest data. Best results are \textbf{bolded}.}\label{table:repeat_rd_gm}\\
\hline
Cortical Regions & GT & RAW & P2S & MPPCA &  BM4D & SWIN \\
\hline
lh-bankssts                 &   1.098 &  2.347 &           2.138 &           2.268 &  2.247 &  \textbf{2.023} \\
lh-caudalanteriorcingulate  &    1.71 &  2.912 &           2.624 &           2.689 &  2.603 &   \textbf{2.51} \\
lh-caudalmiddlefrontal      &   1.422 &   2.41 &           2.322 &           2.343 &  2.323 &  \textbf{2.229} \\
lh-cuneus                   &   1.624 &   2.36 &  \textbf{2.105} &           2.356 &  2.334 &           2.335 \\
lh-entorhinal               &    2.76 &   4.89 &           4.634 &           4.642 &  4.687 &  \textbf{4.591} \\
lh-fusiform                 &  0.9626 &  1.898 &           1.831 &           1.964 &  1.931 &  \textbf{1.782} \\
\hline
\end{longtable}
\begin{longtable}{| p{.34\textwidth} | p{.11\textwidth} | p{.11\textwidth} | p{.11\textwidth} | p{.11\textwidth} | p{.11\textwidth} | p{.11\textwidth} |}
\caption{Coefficient of variation (\%) for AD estimation via no denoising (RAW), P2S, BM4D, MPPCA, and SWIN for gray matter cortical regions in HCP test-retest data. Best results are \textbf{bolded}.}\label{table:repeat_ad_gm}\\
\hline
Cortical Regions & GT & RAW & P2S & MPPCA &  BM4D & SWIN \\
\hline
lh-bankssts                 &   1.121 &           1.997 &           1.934 &           2.172 &  1.976 &  \textbf{1.887} \\
lh-caudalanteriorcingulate  &   1.653 &  \textbf{1.933} &           2.078 &           2.113 &  2.063 &            2.12 \\
lh-caudalmiddlefrontal      &    1.31 &           2.219 &           2.304 &           2.233 &   2.15 &  \textbf{2.003} \\
lh-cuneus                   &   1.311 &           1.914 &           2.043 &           1.831 &  1.834 &    \textbf{1.8} \\
lh-entorhinal               &   2.502 &  \textbf{3.704} &            3.95 &           4.159 &   3.93 &           3.935 \\
lh-fusiform                 &  0.9601 &           1.765 &           1.694 &           1.812 &  1.766 &  \textbf{1.551} \\
\hline
\end{longtable}
\begin{longtable}{| p{.34\textwidth} | p{.11\textwidth} | p{.11\textwidth} | p{.11\textwidth} | p{.11\textwidth} | p{.11\textwidth} | p{.11\textwidth} |}
\caption{Coefficient of variation (\%) for FA estimation via no denoising (RAW), P2S, BM4D, MPPCA, and SWIN for white matter regions in HCP test-retest data. Best results are \textbf{bolded}.}\label{table:repeat_fa_wm}\\
\hline
White Matter Regions & GT & RAW & P2S & MPPCA &  BM4D & SWIN \\
\hline
MCP    &  0.8846 &  \textbf{1.758} &  5.121 &           2.407 &           2.195 &           2.154 \\
PCT    &   1.274 &  \textbf{2.487} &   5.15 &           3.657 &           3.459 &            3.59 \\
GCC    &  0.5437 &  \textbf{1.189} &  4.451 &           1.507 &           1.393 &           1.466 \\
BCC    &  0.6061 &  \textbf{1.857} &  6.761 &           1.903 &           1.918 &           1.964 \\
SCC    &   0.549 &  \textbf{1.441} &  7.807 &           1.638 &           1.671 &           1.769 \\
FX     &   1.508 &  \textbf{2.783} &  8.903 &           3.637 &           3.267 &           3.157 \\
\hline
\end{longtable}
\begin{longtable}{| p{.34\textwidth} | p{.11\textwidth} | p{.11\textwidth} | p{.11\textwidth} | p{.11\textwidth} | p{.11\textwidth} | p{.11\textwidth} |}
\caption{Coefficient of variation (\%) for MD estimation via no denoising (RAW), P2S, BM4D, MPPCA, and SWIN for white matter regions in HCP test-retest data. Best results are \textbf{bolded}.}\label{table:repeat_md_wm}\\
\hline
White Matter Regions & GT & RAW & P2S & MPPCA &  BM4D & SWIN \\
\hline
MCP    &   1.115 &           2.468 &           2.793 &           2.456 &           2.493 &  \textbf{2.333} \\
PCT    &   1.305 &           2.788 &           3.054 &           2.645 &           2.682 &  \textbf{2.468} \\
GCC    &   2.778 &  \textbf{2.009} &            2.93 &           2.213 &           2.202 &           2.177 \\
BCC    &    1.72 &  \textbf{2.435} &           3.093 &           2.617 &           2.622 &           2.457 \\
SCC    &   1.683 &  \textbf{2.268} &           3.598 &           2.533 &           2.527 &           2.573 \\
FX     &   3.043 &           3.625 &            3.94 &           3.715 &           3.565 &  \textbf{3.491} \\
\hline
\end{longtable}
\begin{longtable}{| p{.34\textwidth} | p{.11\textwidth} | p{.11\textwidth} | p{.11\textwidth} | p{.11\textwidth} | p{.11\textwidth} | p{.11\textwidth} |}
\caption{Coefficient of variation (\%) for RD estimation via no denoising (RAW), P2S, BM4D, MPPCA, and SWIN for white matter regions in HCP test-retest data. Best results are \textbf{bolded}.}\label{table:repeat_rd_wm}\\
\hline
White Matter Regions & GT & RAW & P2S & MPPCA &  BM4D & SWIN \\
\hline
MCP    &  1.777 &   3.89 &   \textbf{3.39} &  3.866 &  3.878 &           3.608 \\
PCT    &  1.843 &  4.329 &  \textbf{3.462} &  4.174 &   4.22 &           3.843 \\
GCC    &  2.688 &  4.209 &  \textbf{3.269} &  4.431 &  4.415 &           4.289 \\
BCC    &  3.027 &  6.325 &  \textbf{4.038} &  6.256 &   6.47 &           5.827 \\
SCC    &  3.396 &  6.538 &  \textbf{4.377} &   7.04 &  7.265 &            7.21 \\
FX     &  5.346 &  6.488 &   \textbf{4.07} &  6.619 &  6.684 &           6.392 \\
\hline
\end{longtable}
\begin{longtable}{| p{.34\textwidth} | p{.11\textwidth} | p{.11\textwidth} | p{.11\textwidth} | p{.11\textwidth} | p{.11\textwidth} | p{.11\textwidth} |}
\caption{Coefficient of variation (\%) for AD estimation via no denoising (RAW), P2S, BM4D, MPPCA, and SWIN for white matter regions in HCP test-retest data. Best results are \textbf{bolded}.}\label{table:repeat_ad_wm}\\
\hline
White Matter Regions & GT & RAW & P2S & MPPCA &  BM4D & SWIN \\
\hline
MCP    &  0.9666 &            1.73 &           2.251 &            1.65 &  1.684 &   \textbf{1.58} \\
PCT    &   1.277 &           2.307 &           2.614 &           2.047 &  2.024 &  \textbf{1.865} \\
GCC    &   2.995 &   \textbf{1.74} &           2.762 &           2.047 &    1.9 &           1.909 \\
BCC    &    1.52 &  \textbf{1.627} &           2.767 &            1.74 &  1.663 &           1.678 \\
SCC    &   1.347 &           1.386 &           3.033 &  \textbf{1.381} &  1.385 &           1.381 \\
FX     &   2.441 &           3.258 &           3.824 &           3.366 &  3.061 &  \textbf{2.969} \\
\hline
\end{longtable}
\begin{longtable}{| p{.40\textwidth} | p{.12\textwidth} | p{.12\textwidth} | p{.12\textwidth} | p{.12\textwidth} | p{.12\textwidth} |}
\caption{Coefficient of variation (\%) for NDI estimation via no denoising (RAW), P2S, BM4D, MPPCA, and SWIN for gray matter cortical regions in HCP test-retest data. Best results are \textbf{bolded}.}\label{table:repeat_ndi_gm}\\
\hline
Cortical Regions & RAW & P2S & MPPCA &  BM4D & SWIN \\
\hline
lh-bankssts                 &  4.446 &  4.584 &  4.447 &  4.535 &  \textbf{1.515} \\
lh-caudalanteriorcingulate  &  5.382 &  5.504 &  5.378 &   5.56 &   \textbf{2.84} \\
lh-caudalmiddlefrontal      &  4.747 &  4.848 &  4.746 &   4.79 &  \textbf{2.041} \\
lh-cuneus                   &  4.533 &  4.605 &  4.532 &  4.561 &  \textbf{2.044} \\
lh-entorhinal               &  5.223 &  5.491 &  5.224 &  5.372 &  \textbf{3.385} \\
lh-fusiform                 &  4.042 &  4.234 &  4.044 &  4.012 &  \textbf{1.475} \\
\hline
\end{longtable}
\begin{longtable}{| p{.40\textwidth} | p{.12\textwidth} | p{.12\textwidth} | p{.12\textwidth} | p{.12\textwidth} | p{.12\textwidth} |}
\caption{Coefficient of variation (\%) for ODI estimation via no denoising (RAW), P2S, BM4D, MPPCA, and SWIN for gray matter cortical regions in HCP test-retest data. Best results are \textbf{bolded}.}\label{table:repeat_odi_gm}\\
\hline
Cortical Regions & RAW & P2S & MPPCA &  BM4D & SWIN \\
\hline
lh-bankssts                 &            1.456 &           1.435 &           1.459 &  1.561 &   \textbf{1.218} \\
lh-caudalanteriorcingulate  &            1.894 &           1.931 &           1.897 &  2.087 &   \textbf{1.875} \\
lh-caudalmiddlefrontal      &            2.444 &  \textbf{2.436} &           2.446 &  2.651 &             2.44 \\
lh-cuneus                   &            1.393 &           1.381 &           1.393 &  1.579 &   \textbf{1.286} \\
lh-entorhinal               &             1.61 &           1.622 &           1.624 &  1.801 &   \textbf{1.585} \\
lh-fusiform                 &            1.156 &           1.175 &           1.152 &  1.226 &  \textbf{0.9686} \\
\hline
\end{longtable}
\begin{longtable}{| p{.40\textwidth} | p{.12\textwidth} | p{.12\textwidth} | p{.12\textwidth} | p{.12\textwidth} | p{.12\textwidth} |}
\caption{Coefficient of variation (\%) for FISO estimation via no denoising (RAW), P2S, BM4D, MPPCA, and SWIN for gray matter cortical regions in HCP test-retest data. Best results are \textbf{bolded}.}\label{table:repeat_fiso_gm}\\
\hline
Cortical Regions & RAW & P2S & MPPCA &  BM4D & SWIN \\
\hline
lh-bankssts                 &  20.25 &  20.62 &  20.26 &  22.03 &  \textbf{16.25} \\
lh-caudalanteriorcingulate  &   16.0 &  15.88 &  15.99 &  16.66 &  \textbf{14.27} \\
lh-caudalmiddlefrontal      &  15.85 &  16.57 &  15.86 &  16.71 &   \textbf{12.2} \\
lh-cuneus                   &  12.74 &  13.56 &  12.74 &  13.57 &  \textbf{9.165} \\
lh-entorhinal               &  13.39 &  13.91 &  13.39 &  13.88 &  \textbf{10.67} \\
lh-fusiform                 &  15.35 &  16.14 &  15.35 &  16.23 &   \textbf{10.5} \\
\hline
\end{longtable}
\begin{longtable}{| p{.40\textwidth} | p{.12\textwidth} | p{.12\textwidth} | p{.12\textwidth} | p{.12\textwidth} | p{.12\textwidth} |}
\caption{Coefficient of variation (\%) for NDI estimation via no denoising (RAW), P2S, BM4D, MPPCA, and SWIN for white matter regions in HCP test-retest data. Best results are \textbf{bolded}.}\label{table:repeat_ndi_wm}\\
\hline
White Matter Regions & RAW & P2S & MPPCA &  BM4D & SWIN \\
\hline
MCP    &  1.515 &  1.779 &  1.515 &  1.581 &   \textbf{1.008} \\
PCT    &  2.165 &  2.413 &  2.168 &  2.454 &   \textbf{1.856} \\
GCC    &  2.297 &  2.574 &  2.297 &  2.461 &   \textbf{2.092} \\
BCC    &  2.411 &  2.632 &  2.408 &   2.41 &   \textbf{1.355} \\
SCC    &  1.662 &  1.935 &   1.66 &  1.663 &  \textbf{0.9457} \\
FX     &  3.456 &  3.894 &  3.456 &  3.618 &   \textbf{2.446} \\
\hline
\end{longtable}
\begin{longtable}{| p{.40\textwidth} | p{.12\textwidth} | p{.12\textwidth} | p{.12\textwidth} | p{.12\textwidth} | p{.12\textwidth} |}
\caption{Coefficient of variation (\%) for ODI estimation via no denoising (RAW), P2S, BM4D, MPPCA, and SWIN for white matter regions in HCP test-retest data. Best results are \textbf{bolded}.}\label{table:repeat_odi_wm}\\
\hline
White Matter Regions & RAW & P2S & MPPCA &  BM4D & SWIN \\
\hline
MCP    &           3.362 &           3.657 &           3.388 &           3.313 &  \textbf{3.021} \\
PCT    &           6.411 &           7.023 &            6.45 &           6.495 &  \textbf{5.759} \\
GCC    &            4.48 &           8.522 &           4.374 &  \textbf{4.372} &           4.732 \\
BCC    &           2.954 &           6.584 &  \textbf{2.948} &           3.064 &           3.194 \\
SCC    &           2.629 &           6.068 &   \textbf{2.62} &           2.708 &           3.016 \\
FX     &           4.123 &           6.529 &           4.097 &           3.882 &  \textbf{3.714} \\
\hline
\end{longtable}
\begin{longtable}{| p{.40\textwidth} | p{.12\textwidth} | p{.12\textwidth} | p{.12\textwidth} | p{.12\textwidth} | p{.12\textwidth} |}
\caption{Coefficient of variation (\%) for FISO estimation via no denoising (RAW), P2S, BM4D, MPPCA, and SWIN for white matter regions in HCP test-retest data. Best results are \textbf{bolded}.}\label{table:repeat_fiso_wm}\\
\hline
White Matter Regions & RAW & P2S & MPPCA &  BM4D & SWIN \\
\hline
MCP    &           9.293 &           11.18 &           9.293 &  9.485 &  \textbf{8.199} \\
PCT    &           10.05 &           11.01 &           10.05 &  10.69 &  \textbf{9.466} \\
GCC    &           8.581 &            9.76 &           8.595 &   8.68 &  \textbf{7.502} \\
BCC    &           13.03 &           14.86 &           13.04 &  14.01 &   \textbf{11.4} \\
SCC    &           14.05 &           15.38 &           14.04 &  14.56 &  \textbf{13.58} \\
FX     &            10.7 &           11.71 &            10.7 &  11.24 &   \textbf{9.04} \\
\hline
\end{longtable}